\documentclass[AMA,Times1COL]{WileyNJDv5} 
\usepackage{tikz}
\usetikzlibrary{bayesnet}
\usepackage{makecell}
\usepackage{multicol,multirow}
\usepackage{subcaption}

\articletype{Tutorial in Biostatistics}%

\received{Date Month Year}
\revised{Date Month Year}
\accepted{Date Month Year}
\journal{Journal}
\volume{00}
\copyyear{2025}
\startpage{1}

\begin{document}

\title{Bayesian Regression Analysis with the Drift-Diffusion Model}

\author[1]{Zekai Jin}

\author[2]{Yaakov Stern}

\author[1,3]{Seonjoo Lee}

\authormark{ZEKAI \textsc{et al.}}
\titlemark{BAYESIAN REGRESSION ANALYSIS OF THE DRIFT-DIFFUSION MODEL USING REGDDM}

\address[1]{\orgdiv{Mental Health Data Science}, \orgname{New York State Psychiatric Institute}, \orgaddress{\state{New York}, \country{USA}}}

\address[2]{\orgdiv{Departments of Neurology}, \orgname{Columbia University Irving Medical Center}, \orgaddress{\state{New York}, \country{USA}}}

\address[3]{\orgdiv{Department of Biostatistics}, \orgname{Columbia University Irving Medical Center}, \orgaddress{\state{New York}, \country{USA}}}

\corres{Seonjoo Lee \email{sl3670@cumc.columbia.edu}}



\abstract[Abstract]{
  The Drift-Diffusion Model (DDM) is widely used in neuropsychological studies to understand the decision process by incorporating both reaction times and subjects' responses. Various models have been developed to estimate DDM parameters, with some employing Bayesian inference. However, when examining associations between phenotypes of interest and DDM parameters, most studies adopt a two-step approach: first estimating DDM parameters, then applying a separate statistical model to the estimated values. Despite the potential for bias, this practice remains common,  primarily due to researchers' unfamiliarity with Bayesian modeling. To address this issue, this tutorial presents the implementations and advantages of fitting a unified Bayesian hierarchical regression model that integrates trial-level drift-diffusion modeling and subject-level regression between DDM parameters and other variables. The R package RegDDM, developed and demonstrated in this tutorial, facilitates this integrated modeling approach.
}


\jnlcitation{\cname{%
\author{Jin Z},
\author{Stern Y}, and
\author{Lee S}}.
\ctitle{Regression analysis with DDM}
\cjournal{\it J Comput Phys.???} \cvol{2021;00(00):1--18???}.}

\maketitle

\renewcommand\thefootnote{}
\footnotetext{\textbf{Abbreviations:} DDM, Drift-Diffusion Model; PDR: Performance Degradation Rate}

\renewcommand\thefootnote{\fnsymbol{footnote}}
\setcounter{footnote}{1}

\section{Introduction}\label{Introduction}
In cognitive psychology and cognitive neuroscience, many researchers use binary decision-making tasks to study human behavior \cite{kirchner1958age, meyer1971facilitation, bonnelle2015characterization}. During these tasks, subjects are presented with certain types of information and asked to make binary decisions based on the information provided. The response and reaction time are recorded for downstream analysis. Traditionally, the rate of specific response and the reaction time were analyzed separately. While simple, this approach fails to account for other aspects of the decision-making process, such as the trade-off between speed and accuracy. To address this limitation, the Drift-Diffusion Model (DDM) was introduced \cite{ratcliff1978theory}. The DDM conceptualizes binary decision-making as an evidence-accumulating process governed by three DDM parameters: non-decision time, representing the gap between the stimulus and initial accumulation; drift rate, indicating the speed of evidence accumulation; and threshold, determining the amount of evidence required to reach a decision. Each of these DDM parameters reflect different aspects of cognitive processing \cite{lerche2016model} and has gained widespread use in many applications \cite{mads2023computational, gold2007neural, forstmann2016sequantial}.

Since its introduction, numerous improvements have been proposed to address limitations of the original DDM. First, to reconcile discrepancy between observed and predicted reaction time \cite{anderson1960modification, laming1968information}, a full-DDM was proposed with additional parameters to account for the inter-trial variability for DDM parameters \cite{ratcliff2004comparison}, and recently generalized to account for arbitrary distributions for such variability \cite{shinn2020flexible}. Second, to improve computation efficiency, an approximate method was introduced to compute the probability density function of reaction time distribution, which is widely used in many packages fitting DDM \cite{navarro2009fast}. Furthermore, to improve the estimation of DDM parameters, HDDM adopted Bayesian hierarchical modeling that allows pooling of information and thus becomes more robust against outliers \cite{wiecki2013hddm}.

\begin{figure}[h]
    \centering
    \includegraphics[width=0.8\linewidth]{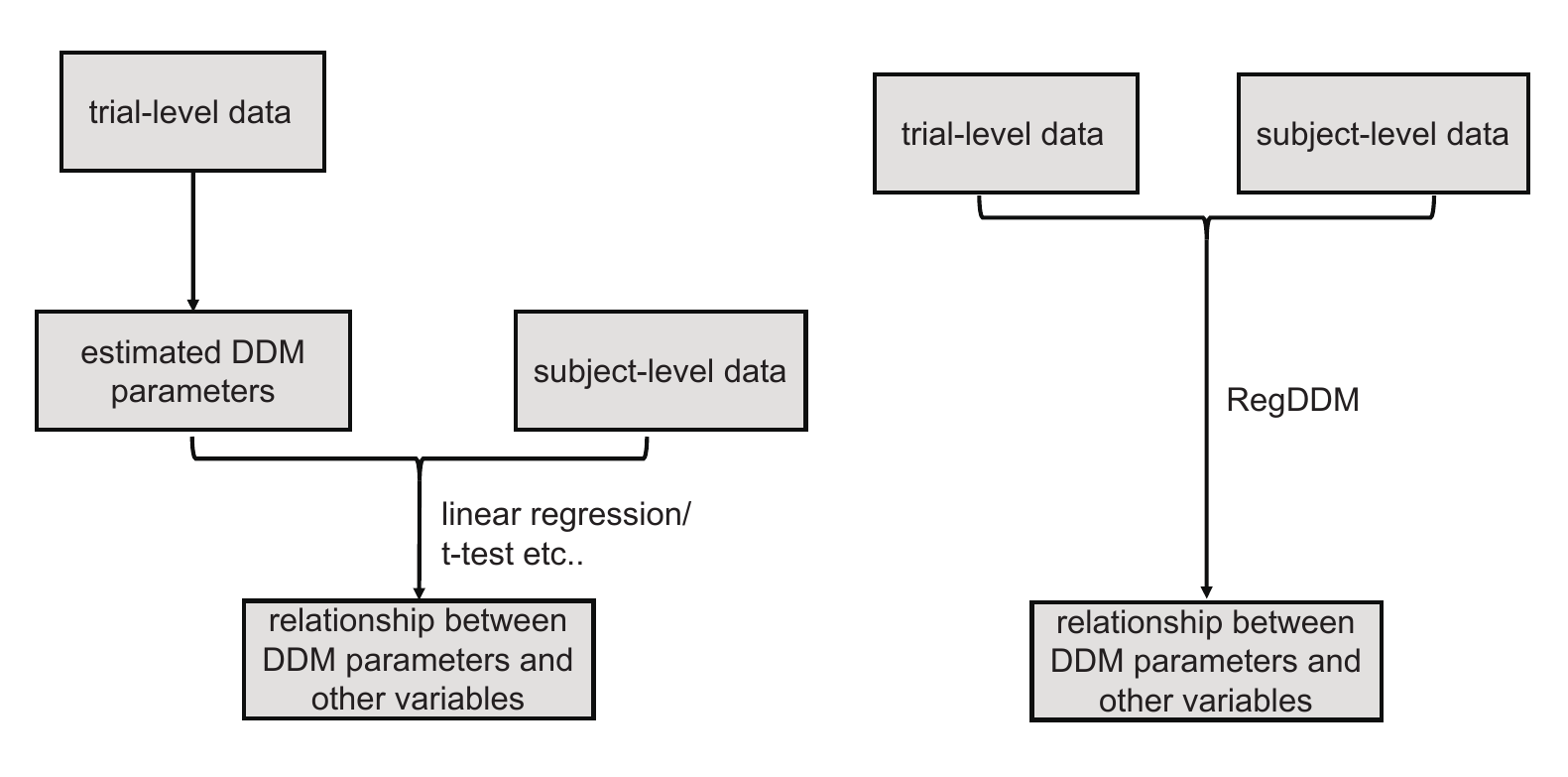}
    \caption{Workflow comparison between the two-step approach and fitting a single model with RegDDM}
    \label{workflow_comparison}
\end{figure}

Recently, more neuroscience studies employ DDM to link the decision process to participant’s demographic, clinical, and other biological characteristics \cite{mads2023computational, schmiedek2007individual, white2010using}. To examine whether outcome variables are associated with specific DDM parameters, most studies adopt a two-step approach. (Figure \ref{workflow_comparison}) In the first step, DDM parameters are estimated for each subject using only trial-level data, including response, reaction time, and the trial conditions. In the second step, a separate statistical analysis, such as linear regressions or t-tests, is performed to assess the association between the estimated DDM parameters and other variables of interest. However, this two-step approach has two major limitations. First, regardless of the method used for DDM parameter estimation, inferences drawn in the second step may be inaccurate due to unaccounted measurement error in the estimated DDM parameters and also failure to integrate information across different hierarchical levels. Second, the mixing Bayesian and Frequentist methods in current practice complicates the interpretationn of results and undermines methodological coherence. 

One way to address the limitations of two-step analysis is to build a single model and employ proper statistical inference. With the development of Markov Chain Monte Carlo (MCMC) methods and the increasing computing power, there has been growing interest in the use of Bayesian hierarchical modeling when analyzing binary choice data with the DDM \cite{wiecki2013hddm, nunez2024tutorial}. Thus, we focus on Bayesian hierarchical modeling in this tutorial. Compared to other approaches, Bayesian hierarchical modeling offers several advantages. First, it allows the incorporation of prior knowledge, which is particularly useful when historical data or expert judgment are available. Second, it naturally handles missing data as latent variables, whereas Frequentist approaches often imputation or deletion. Third, Bayesian hierarchical modeling is highly flexible and can accommodate complex model structures. Additionally, the interpretation of Bayesian posterior distributions is often more intuitive than Frequentist estimates \cite{merkle2011comparison, bernardo2009bayesian, gelman1995bayesian}. Various tools have been developed to perform Bayesian statistical analysis using MCMC. These packages enables users to customize model structures and provide built-in tools for model fitting and diagnostics \cite{rstan, abril2023pymc}.

The objective of this tutorial is to provide detailed step-by-step guidance on fitting a single Bayesian regression model with the DDM. In Section \ref{Example_study}, we introduce the Letter-Sternberg (LS) task dataset from the Cognitive Reserve study \cite{steffener2009impact}. Section \ref{Methods} presents detailed methods for the Bayesian regression model and the R package RegDDM , which we developed to implement it. Section \ref{Application} demonstrated the results from both the integrated and two-step approaches.  Section \ref{Simulation} compared the performance of the RegDDM through simulation and evaluates the influence of sample size on parameter estimation. Finally, we conclude with several advanced topics in Section \ref{Additional}, such as building complex models and handling missing data.  

\section{Example Study}\label{Example_study}
The Cognitive Reserve study \citep{steffener2009impact, coors2024brain} is an ongoing longitudinal study  to identify neural implementations of cognitive reserve. In the study, each participant completed the LS task \citep{sternberg1966high}. During the task, the subject was first presented with a set of uppercase letters for 3 seconds. After a 7-second retention interval, a single uppercase letter was shown as a probe. Upon seeing the probe, the subject was instructed to decide whether the probe was one of the letters presented at the beginning of the trial as quickly as possible and to press the corresponding button. The subject's response (correct = 1, incorrect = 0) and response time for each trial were then recorded. The number of letters presented in the initial set was either 1, 3, or 6, indicating different memory loads. Trials were run in batches of 10, and the total number of trials per subject ranged from 30 to 180. Some trials were excluded if the response time was less than 0.1 second or greater than 3 seconds. 

In addition to task performance data, demographic variables such as age, gender, and Intelligence Quotient (IQ), measured using the American National Adult Reading Test (NART) \citep{grober1991development} were collected. For this illustration, we included 49 young participants aged 20-30. We also calculated the summary statistic for the proportion of correct answers (Accuracy) and average response time (Latency) for each subject. The summary statistics are presented in Table \ref{summarystat}. 

\begin{table}[h]
\centering
\begin{tabular}{lll}
\hline
Variable & & Mean(SD)/n(\%) \\
\hline
Age, years  & & 26.35(3.03) \\
Education, years   &&  15.71(1.94) \\
Gender   &  & \\
&F &  33(67\%) \\
&M &  16(33\%) \\
Race & &   \\
&Asian  & 3(6.1\%)  \\
&Black  &  10(20\%)  \\
&White  &  8(57\%) \\
&Other  &  28(16\%)  \\
NART-IQ  & & 117(7) \\
Accuracy  & & 0.94(0.09) \\
Latency, second  & & 1.14(0.23) \\
\hline
\end{tabular}
\caption{Summary statistics for selected Cognitive Reserve study dataset. Statistics are Mean(SD) for continuous variables and n(\%) for factor variables}
\label{summarystat}
\end{table}

The dataset consists of two separate data frames at different levels. The first is a subject-level data frame containing information such as id, age and NART-IQ. The second is the trial-level data frame with variables including subject id, task difficulty, response and reaction time for each trial. We note that the original data used here cannot be uploaded in the public domain due to the data use agreement. The dataset used in this tutorial is a synthetic version of the original using R packages synthpop \cite{synthpop} and rtdists \cite{rtdists}. This synthetic dataset can be assessed through our R package RegDDM. Table \ref{example_data1} and \ref{example_data2} showed the first 3 rows of each synthetic data frame. Qualified researchers interested in analyzing the original dataset may request access by contacting the authors.

\begin{table}[h]
\centering
\begin{tabular}{cccccc}
\hline
id & iq & age & gender & race & education \\
\hline
1 & 112.8& 22& F&White &14 \\
2 & 114.3& 22& F&White &16 \\
3 & 117.0& 22& F&Black &13 \\
\multicolumn{6}{c}{\ldots}  \\
\hline
\end{tabular}
\caption{Example subject-level data frame.}
\label{example_data1}
\end{table}
\begin{table}[h]
\centering
\begin{tabular}{cccc}
\hline
id & memload & response & rt \\
\hline
1 & 1 & 1 & 1.18 \\
1 & 1 & 1 & 2.22 \\
1 & 6 & 1 & 4.42 \\
\multicolumn{4}{c}{\ldots}  \\
\hline
\end{tabular}
\caption{Example trial-level data frame. memload is the memory load (number of letters) of the trial. response is the subject's response of the trial (Correct = 1) and rt is the response time in seconds.}
\label{example_data2}
\end{table}

\section{Methods}\label{Methods}
\subsection[ddm]{Drift-diffusion model}
\begin{figure}[h]
    \centering
    \includegraphics[width=0.8\linewidth]{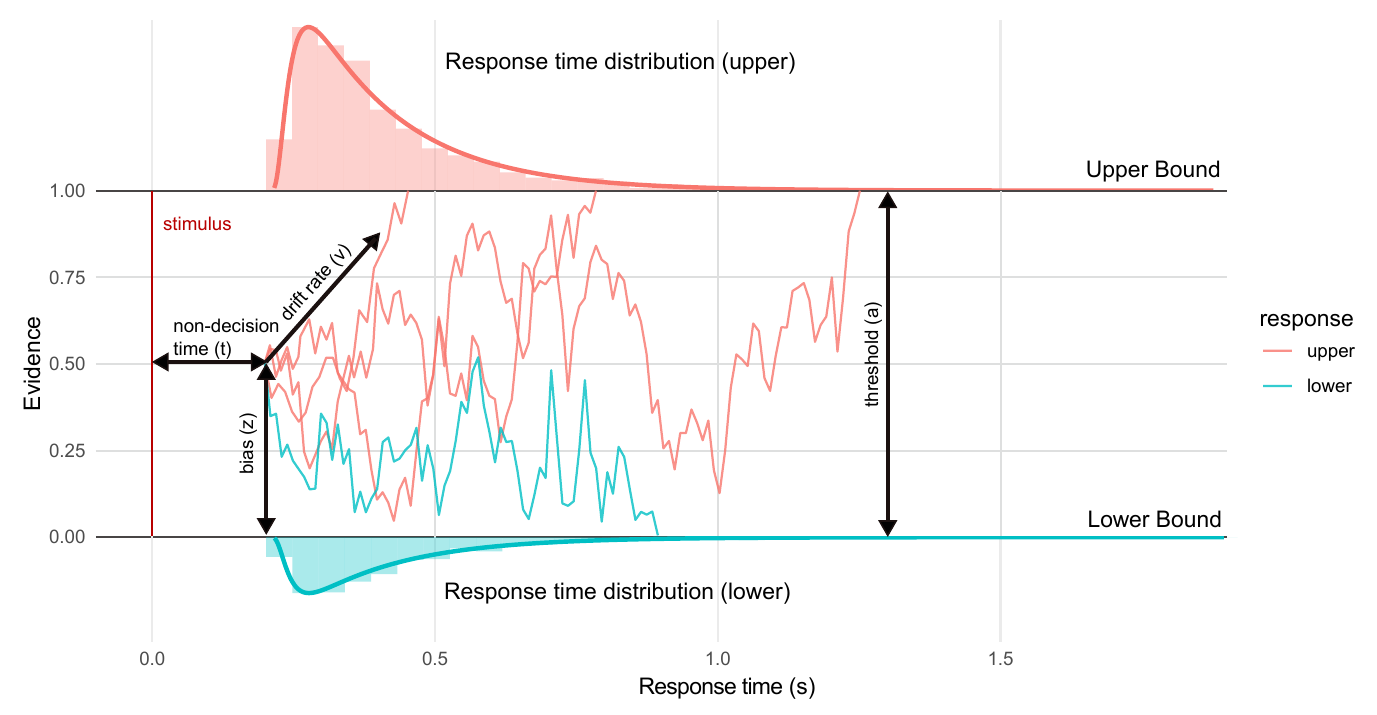}
    \caption{Demonstration of the four parameter drift-diffusion model}
    \label{ddm_figure}
\end{figure}

In neuropsychological studies, the DDM is widely used to model subjects' behavior during decision-making tasks, including the LS task. The DDM considers binary decision as an evidence accumulation process (Figure \ref{ddm_figure}). For each trial, the subject starts accumulating information shortly after the stimulus in a noisy, Brownian motion-like way. Once the accumulated evidence reaches one of the two decision boundaries, the subject makes the decision accordingly. The boundary determines the binary response, while the time to reach the boundary determines the reaction time. Although there are more complex variants of DDM, four-parameter models require fewer trials for robust parameter estimation\cite{lerche2016model, shinn2020flexible}. This simpler model is also widely used across studies and is therefore adopted in this paper.

The four parameters are drift rate $v$, representing the speed of evidence accumulation; non-decision time $t$, representing the delays between stimulus and the start of drift-diffusion process; threshold $a$, representing the total evidence needed for decision making, and bias $z$, which controls the starting point of the decision. Each of these four parameters captures different aspects of the cognitive process, and our objective is to study the relationship between these DDM parameters and other variables of interest. 

To formulate the model, denote the total number of subjects as $N$ and index the subjects using $i = 1,2,\ldots,N$. Denote the number of trials for subject $i$ as $n_i$ and index trials using $j=1,2,\ldots,n_i$. For trial $j$ of subject $i$, denote the four DDM parameters as $v_{i,j}$, $a_{i,j}$, $t_{i,j}$ and $z_{i,j}$. Denote the reaction time as $r_{i,j}$ and the response as $s_{i,j}$  (Upper$=1$, Lower$=0$). As formulated by DDM, if $s_{i,j} = 1$, $r_{i,j}$ is assumed to follow a Wiener First Passage Time (WFPT) distribution:
\[r_{i,j}\sim WFPT(a_{i,j},t_{i,j},z_{i,j},v_{i,j})\]
The PDF of WFPT distribution is defined as follows:
\[
WFPT(x; a, t, z, v) = \frac{a^3}{(x-t)^{\frac{3}{2}}} \exp\left(-vaz - \frac{v^2 (x-t)}{2}\right) \sum_{k=-\infty}^\infty (2k+z) \phi\left(\frac{2ka+z}{\sqrt{x-t}}\right)
\]
For trials with lower responses, it is equivalent to upper responses with $-v_{i,j}$ drift rate and $1-z_{i,j}$ bias. Thus, 
\begin{align}
r_{i,j} \sim &
    \begin{cases}
        WFPT(a_{i,j}, t_{i,j},z_{i,j},v_{i,j}), & \text{if }s_{i,j} = 1\\
        WFPT(a_{i,j}, t_{i,j},1-z_{i,j},-v_{i,j}), & \text{if }s_{i,j} = 0\\
    \end{cases}
    \label{eq1}
\end{align}

In the LS task, we focused on analyzing drift rate due to its direct connection with subjects' performance. Typically, when the trials are more difficult, the subjects' accuracy will decrease and response time will increase, corresponding to smaller drift rates. To model such effect, we assume a deterministic linear relationship between memory load and drift rate specified by equation \ref{eq2}, where ${x}_{i,j}$ is the memory load of trial $j$, $v_{0,i}$ is subject $i$'s baseline drift rate and $v_{x,i}$ measures subject $i$'s performance degradation when difficulty increases:
\begin{equation}
    v_{i,j} = v_{0,i}+{x}_{i,j}{v}_{x,i}
    \label{eq2}
\end{equation}

In this tutorial, we use ${v}_{x,i}$ as a measure for the performance degradation rate (PDR) of a subject.  For most subjects, ${v}_{x,i}$ is negative. The higher ${v}_{x,i}$ is, the slower their performance degrades as the difficulty of the trial increases. For the three other DDM parameters, we assume that they are not influenced by the difficulty of the trial to reduce model complexity. Thus, the DDM parameters of the trial is the same as the baseline DDM parameters of the subject.
\begin{align}
    a_{i,j} &= a_{0,i}\nonumber \\
    t_{i,j} &= t_{0,i}\nonumber\\
    z_{i,j} &= z_{0,i}
    \label{eq3}
\end{align}

For computation efficiency, trial-level DDM parameters $a_{i,j}$, $t_{i,j}$, $z_{i,j}$ and $v_{i,j}$ are deterministic. Only subject-level DDM parameters $a_{0,i}$, $t_{0,i}$, $z_{0,i}$, $v_{0,i}$ and $v_{x,i}$ are stochastic. To improve estimation, RegDDM adopts a hierarchical approach similar to HDDM \cite{wiecki2013hddm}. Each of the five subject-level DDM parameters is assumed to follow a normal distribution with their respective mean and standard deviation, as formulated in Equation \ref{eq4}. This hierarchical approach in estimating subject-level DDM parameters allows pooling of information, thus improves stability and accuracy especially for smaller sample sizes. 
\begin{align}
    a_{0,i} &\sim N(\mu_{a_0},\sigma_{a_0}^2)\nonumber\\
    t_{0,i} &\sim N(\mu_{t_0},\sigma_{t_0}^2)\nonumber\\
    z_{0,i} &\sim N(\mu_{z_0},\sigma_{z_0}^2)\nonumber\\
    v_{0,i} &\sim N(\mu_{v_0},\sigma_{v_0}^2)\nonumber\\
    v_{x,i} &\sim N(\mu_{v_x},\sigma_{v_x}^2)
    \label{eq4}
\end{align}

\subsection[RegDDM]{Regression Drift-Diffusion Model(RegDDM)}\label{regddm_intro}
\begin{figure}[h]
    \centering
    \begin{subfigure}[t]{0.45\textwidth}
    \centering
    \begin{tikzpicture}


      \node[latent](vx){${v_{x,i}}$};
      \node[latent, below = of vx](v0){$v_{0,i}$};
      \node[latent, below = of v0](z0){$z_{0,i}$};
      \node[latent, below = of z0](t0){$t_{0,i}$};   
      \node[latent, below = of t0](a0){$a_{0,i}$};
      
      \node[obs, right = of vx, xshift = 1.8cm] (y) {$\text{iq}_{i}$};

      \node[det, right = of v0] (v) {$v_{i,j}$};
      \node[det, right = of z0] (z) {$z_{i,j}$};
      \node[det, right = of t0] (t) {$t_{i,j}$};
      \node[det, right = of a0] (a) {$a_{i,j}$};
      
      \node[obs, right = of t] (rt) {$r_{i,j}$};
      \node[obs, above = of rt] (r) {$s_{i,j}$};
      
      \node[obs, right = of v] (x) {${x_{i,j}}$};
    

      \edge{v0} {y}; %
      \edge[color = red]{vx} {y}; %

      \edge{vx} {v}; %
      \edge{a0} {a}; %
      \edge{z0} {z}; %
      \edge{t0} {t}; %
      \edge{v0} {v}; %

      \edge{x} {v};
    
      \edge{a} {rt};
      \edge{z} {rt};
      \edge{t} {rt};
      \edge{v} {rt};
      \edge{r} {rt};
      
      \plate {j} {(x)(a)(z)(t)(v)(rt)} {Trial $j$} ;
      \plate {i} {(x)(a)(z)(t)(v)(a0)(rt)(vx)(y)(j.south east)(j.south west)} {Subject $i$};
    \end{tikzpicture}
    \caption{Model used in analysis 1}
    \label{model_overviewA}
    \end{subfigure}
    \hspace{1cm}
    \begin{subfigure}[t]{0.45\textwidth}
        \centering    
    \begin{tikzpicture}

 
      \node[latent](vx){${v_{x,i}}$};
      \node[latent, below = of vx](v0){$v_{0,i}$};
      \node[latent, below = of v0](z0){$z_{0,i}$};
      \node[latent, below = of z0](t0){$t_{0,i}$};   
      \node[latent, below = of t0](a0){$a_{0,i}$};
      
      \node[obs, right = of vx, xshift = 1.8cm] (y) {$\text{gender}_{i}$};

      \node[det, right = of v0] (v) {$v_{i,j}$};
      \node[det, right = of z0] (z) {$z_{i,j}$};
      \node[det, right = of t0] (t) {$t_{i,j}$};
      \node[det, right = of a0] (a) {$a_{i,j}$};
      
      \node[obs, right = of t] (rt) {$r_{i,j}$};
      \node[obs, above = of rt] (r) {$s_{i,j}$};
      
      \node[obs, right = of v] (x) {${x_{i,j}}$};
    

      \edge[color = red]{y} {vx}; %

      \edge{vx} {v}; %
      \edge{a0} {a}; %
      \edge{z0} {z}; %
      \edge{t0} {t}; %
      \edge{v0} {v}; %

      \edge{x} {v};
    
      \edge{a} {rt};
      \edge{z} {rt};
      \edge{t} {rt};
      \edge{v} {rt};
      \edge{r} {rt};
      
      \plate {j} {(x)(a)(z)(t)(v)(rt)} {Trial $j$} ;
      \plate {i} {(x)(a)(z)(t)(v)(a0)(rt)(vx)(y)(j.south east)(j.south west)} {Subject $i$};
    \end{tikzpicture}
    \caption{Model used in analysis 2}
    \label{model_overviewB}
    \end{subfigure}

    \caption{Overview of example models used by RegDDM in analysis 1 (a) and analysis 2 (b). Circles represent stochastic nodes. Rhombuses represent fully determined nodes. Nodes in gray are observed variables and nodes in white are latent parameters. Arrows in red are the primary effects of interest. For simplicity purposes, nodes of covariates, regression coefficients, group means and variances of DDM parameters are not displayed.}
    \label{model_overview}
\end{figure}
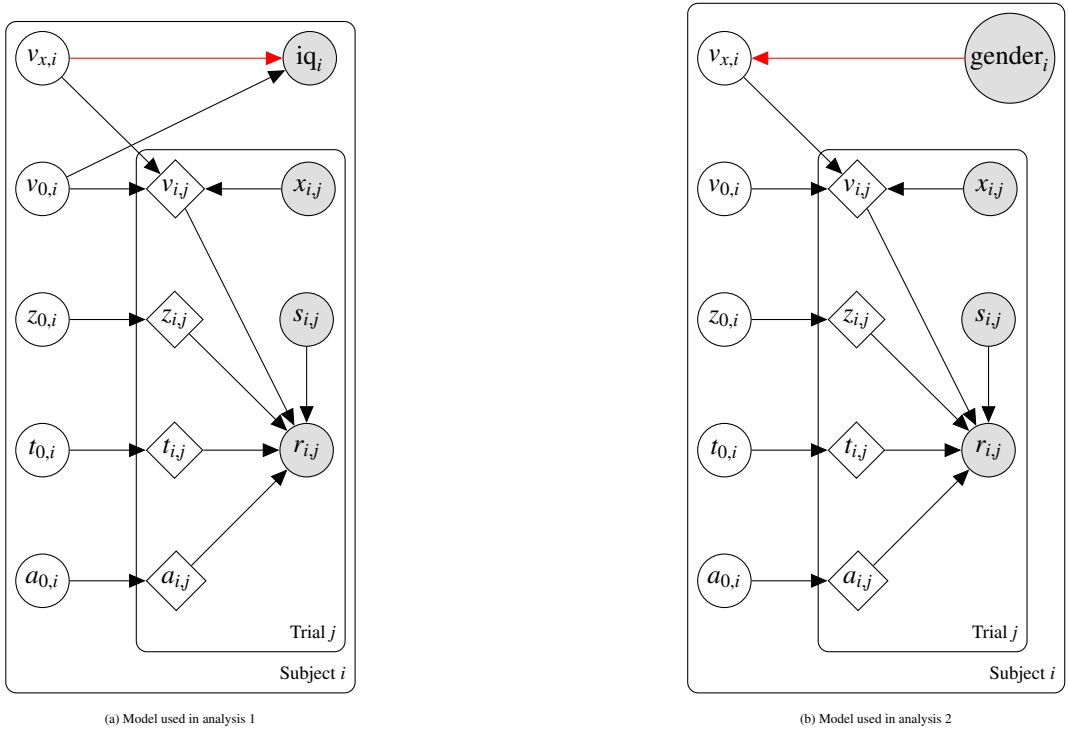

The primary focus of this tutorial is the relationship between the PDR of a subject and other variables. In this tutorial, we focused on two example analyses. The first analysis studied whether the NART-IQ of the subject can be predicted by the PDR, adjusting for baseline drift rate, age, and education level. Specifically, the linear model formulated by Equation \ref{eq5} is used for the first analysis.
\begin{equation}
    \text{iq}_i \sim N(\beta_0 + \beta_{v_{x}}v_{x,i} + \beta_{v_0}v_{0,i} + \beta_{age}\text{age}_{i} + \beta_{education}\text{education}_i,\sigma^2)
    \label{eq5}
\end{equation}
In traditional two-step approach, such analysis is performed by fitting a linear regression model with estimated $v_{0,i}$ and $v_{x,i}$. In RegDDM, we build a single model by stacking another layer of regression model on top of the aforementioned DDM. Figure \ref{model_overviewA} illustrates the final model for analysis 1.

The second analysis studied whether PDR differs by gender. In the two-step approach, such analysis is performed with two-sample t-tests using estimated $v_{x,i}$, or equivalently, with a linear regression model specified by Equation \ref{eq6}.
\begin{equation}
v_{x,i} \sim N(\beta_0+\beta_{gender}\text{gender}_i, \sigma_{v,x}^2)
    \label{eq6}
\end{equation}

This represents the scenario when subject-level DDM parameters are used as the outcome. In RegDDM, we build a single model by replacing Equation \ref{eq4} with Equation \ref{eq6}. Figure \ref{model_overviewB} illustrates the final model for analysis 2.

%

\subsection[bayesian]{Bayesian hierarchical modeling}
In this tutorial, we applied Bayesian hierarchical modeling due to advantages mentioned in the Introduction. In Bayesian statistics, the parameters are considered stochastic and follow certain distributions. Denote the vector of all parameters in the model as $\boldsymbol{\theta}$ and the sample as $\boldsymbol{X}$. Specifically, $p(\theta)$ is the prior distribution, which is the distribution of parameters when no sample is observed, representing the information from previous research. $p(\boldsymbol{X}|\boldsymbol{\theta})$ is the likelihood function specified by the model. The objective of Bayesian inference is to obtain the distribution of $\boldsymbol{\theta}$ conditional on $\boldsymbol{X}$, namely the posterior distribution, using Bayes' rule:

\[p(\boldsymbol{\theta}|\boldsymbol{X})\propto p(\theta)p(\boldsymbol{X}|\boldsymbol{\theta})\]

For DDM, there have been empirical distributions for each of the four parameters \cite{matzke2009psychological}. Thus, as a default setting in RegDDM, we apply the following priors over the group mean and standard deviation of the four subject-level intercepts, which is inherited from previous literature \cite{wiecki2013hddm}.
\begin{align}
    &\mu_{a_0}\sim G(1.125, 0.75)\nonumber\\
    &\mu_{t_0}\sim G(0.08, 0.2)\nonumber\\
    &\mu_{z_0}\sim N(0.5, 0.5)\nonumber\\
    &\mu_{v_0}\sim N(2, 3)\nonumber\\
    &\sigma_{a_0}\sim HN(0, 0.1)\nonumber\\
    &\sigma_{t_0}\sim HN(0, 1)\nonumber\\
    &\sigma_{z_0}\sim HN(0, 0.05)\nonumber\\
    &\sigma_{v_0}\sim HN(0, 2)
    \label{eq7}
\end{align}

Here, $G$ represents a Gamma distribution parameterized by shape and rate. $N$ represents a normal distribution with mean and standard deviation. $HN$ represents a half-normal distribution parameterized by the mean and standard deviation of the corresponding normal distribution. All other parameters of the model will apply non-informative priors since no prior knowledge is available.

The prior distribution in Equation \ref{eq7} should be interpreted as the prior belief of the four DDM parameters when $x_{i,j}=0$. The validity of such interpretation depends heavily on the scale of trial-level variables as well as the experiment setup. In the most extreme scenario, if the trial-level variable is offset by a large value, the intercept will also be offset by a large margin, which may violate the conditions where the prior distributions were obtained. In that scenario, it is recommended to scale the trial-level variables or use non-informative priors for these intercepts.

Once we specify the prior and likelihood, the posterior distribution is determined. However, an analytical expression is typically impossible to obtain. One strategy is to sample from the posterior distribution using MCMC. There have been various packages employing MCMC. In this tutorial, we use RStan, which uses No-U-turn sampler, an extension to the Hamiltonian Monte Carlo sampler \cite{rstan,matthew2014nouturn}. After the sampling process, the posterior mean and 95\% credible interval are estimated using the sample, which can be used for downstream statistical inference.

To facilitate the use of RegDDM, we developed an R package named RegDDM, which can be installed through CRAN using the following R command.
\begin{lstlisting}[basicstyle=\fontsize{8}{10}\selectfont\ttfamily]
install.packages('RegDDM')
\end{lstlisting}
The package takes the data in a user-friendly way, automatically generates the model code, and fits the model using RStan. It also contains the example LS task dataset used in this tutorial. 

\subsection[twostep]{Two step approach}
This tutorial applied the two step approach for comparison purposes. For the first step, we fit a regular DDM using RegDDM without the regression part in Section \ref{regddm_intro}. The posterior means of subject-level DDM parameters are extracted as estimated values. Then, as a second step, we treat the estimated values as fixed, and fit another linear regression model in the Frequentist framework to study the effects of interest. This procedure aligns with other publications, thus serves as a reference for the current implementation.

\subsection[computing]{Computing environment}
To demonstrate and test our package, we fit a set of models in the Tutorial, Simulation results, and Application sections. Computation in Section \ref{Application} was performed using an Intel Core i7-14700k CPU and computation in Section \ref{Simulation} was performed using an AMD EPYC 9654 CPU. Each chain takes at least 1 CPU core for maximum performance. Default 500 warm-up iterations and 1000 total iterations were run. RegDDM version 1.1 was used, which was built under RStan version 2.32.7 and R version 4.5.1. The R code to replicate all results in this tutorial can be found in the Supporting Information.

\section{Analysis results}\label{Application}
\subsection[Application1]{Analysis 1: IQ prediction with PDR}
In the first analysis, we studied whether the NART-IQ of the subject can be predicted by the PDR, adjusting for covariates. The following code is used to perform the analysis specified by Equation \ref{eq5} in the RegDDM framework: 
\begin{lstlisting}[basicstyle=\fontsize{8}{10}\selectfont\ttfamily]
library(RegDDM)
data(regddm_data)
model1 = list(
  v ~ memload,
  iq ~ v_0 + v_memload + age + education
)
regddm(regddm_data$data1, regddm_data$data2, model1)
\end{lstlisting}

The code first loaded the package and the LS task dataset, then defined the model using a list, and finally drew samples from the model. The computation takes around 20 minutes. The output of the model is as follows.

\begin{lstlisting}[basicstyle=\fontsize{8}{10}\selectfont\ttfamily]
RegDDM Model Summary
Number of subjects: 49
Number of trials: 6032
Model:
  v ~ memload
  iq ~ v_0 + v_memload + age + education
Family: gaussian
Sampling: 4 chains, 500 warmups and 1000 iterations were used. Longest elapsed time is 1272 s.

Regression coefficients:
        variable     mean     sd     2.5%   97.5% n_eff  Rhat
1         beta_0 112.2063 11.997   88.766 134.809  1168 1.005
2       beta_v_0  -3.7177  2.034   -7.646   0.389  2428 0.999
3 beta_v_memload -55.7170 28.413 -117.613  -6.666   861 1.006
4       beta_age   0.1345  0.312   -0.462   0.751  1460 1.004
5 beta_education  -0.0152  0.591   -1.175   1.160  2323 0.999
6          sigma   6.7599  0.793    5.399   8.463  1676 1.001
Maximum R-hat: 1.007
\end{lstlisting}

\begin{figure}[h]
    \centering
    \includegraphics[width=0.5\linewidth]{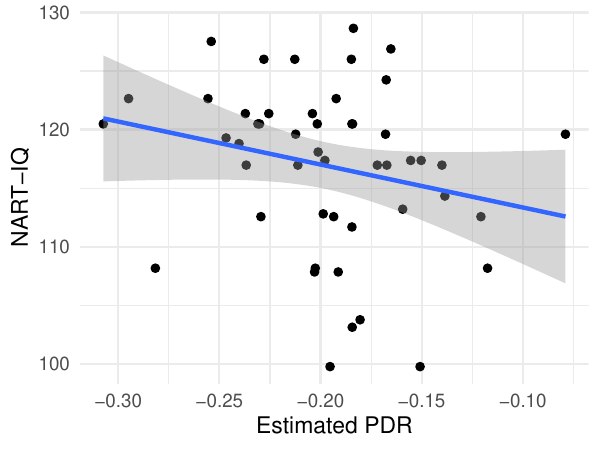}
    \caption{scatter plot of estimated PDR (posterior mean of $\boldsymbol{v_x}$) and NART-IQ.}
    \label{example_scatterplot}
\end{figure}

Counterintuitively, the result indicates that while controlling for other variables, the higher the PDR, the lower the IQ. In other words, subjects who are less susceptible to increased memory load will have lower IQs. To further validate the result, we prepared a scatter plot showing the pairwise relationship between the estimated cognitive reserve and IQ (Figure \ref{example_scatterplot}). From the plot, it is clear that all subjects' performance degrades as the difficulty increases. Compared to subjects who are more susceptible to increased memory loads, those who are less susceptible tend to have lower IQ scores. 

We also performed the analysis using two-step approach. The estimated effect of cognitive reserve on IQ is $-53.74$ (t$(44)=-2.22,p=0.03$), slightly lower than the RegDDM estimates. We found that underestimating effects is common for two-step approaches, which is further analyzed in later Chapters.

\subsection[Application2]{Analysis 2: Gender difference of PDR}
In our second analysis, we compare the PDR between males and females using Equation \ref{eq6}. The following code is used to perform the analysis in the RegDDM framework: 

\begin{lstlisting}[basicstyle=\fontsize{8}{10}\selectfont\ttfamily]
model2 = list(
  v ~ memload,
  v_memload ~ gender
)
regddm(regddm_data$data1, regddm_data$data2, model2)
\end{lstlisting}

Contrary to the first analysis, DDM parameter is used as the outcome in the second analysis. The computation takes around 15 minutes, and the output of the model is displayed as follows:

\begin{lstlisting}[basicstyle=\fontsize{8}{10}\selectfont\ttfamily]
RegDDM Model Summary
Number of subjects: 49
Number of trials: 6032
Model:
  v ~ memload
  v_memload ~ gender
Family: gaussian
Sampling: 4 chains, 500 warmups and 1000 iterations were used. Longest elapsed time is 846 s.

Regression coefficients:
      variable     mean     sd    2.5%   97.5% n_eff Rhat
1       beta_0 -0.19395 0.0150 -0.2228 -0.1647  1246    1
2 beta_genderM -0.00836 0.0243 -0.0532  0.0400  1459    1
3        sigma  0.06200 0.0113  0.0422  0.0855   782    1
Maximum R-hat: 1.008
\end{lstlisting}

According to the result, the average PDR in female subjects is about $-0.19$. Compared to female subjects, male has a negligible smaller PDR around $0.01$ and the standard deviation of the error term is about $0.06$. Based on the 95\% credible interval of beta\_genderM, there is no strong evidence suggesting gender differences in PDR.

For two-step approach, we performed a two-sample t-test to test the difference of estimated cognitive reserve between males and females. The result yields a p-value of 0.68, showing nonsignificant results.

\newpage
\section{Simulation results}\label{Simulation}
\subsection[Simulation1]{Simulation 1: Comparison with two-step approaches}
Two better compare RegDDM and two step approach, we performed a set of simulations. Our first simulation compared the performance between fitting a single model using RegDDM and the two-step approach. It assumes the following relationship between a subject's baseline drift rate $v_{0,i}$ and predictor $u_i$:
\begin{align*}
    u_i&\sim N(0,0.5^2)\\
    v_{0,i}&\sim N(1.5+u_i,0.5^2)
\end{align*}

The full detail of the generation setup can be found in Appendix \ref{Appendix1}. After generating the subject-level DDM parameters, the response and reaction time in the trial-level are simulated using R package rtdists \cite{rtdists}. Then, we implemented the model specified by Equation \ref{eq8} using RegDDM and two-step approach respectively, and compared their performance in estimating the effect of $u_i$ on $v_{0,i}$ .
\begin{equation}
    v_{0,i} \sim N(\beta_0+\beta_uu_i,\sigma^2_{v,0})
    \label{eq8}
\end{equation}

\begin{figure}[h]
    \centering
    \includegraphics[width=0.9\linewidth]{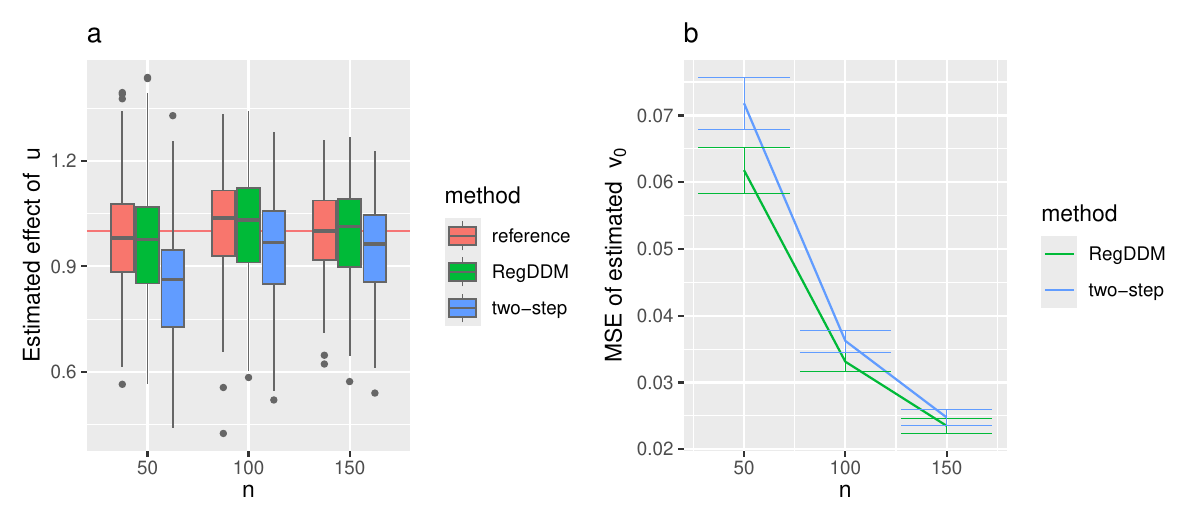}
    \caption{Comparison between methods when $v_0$ is treated as the outcome. (a) Comparison between reference model, two-step approach and RegDDM in estimating the effect of $u$ on $v_0$. Red line represents true effect. (b) MSE of estimated $v_0$ between RegDDM and two-step Approach.}
    \label{sim1_1}
\end{figure}

We also fit another reference linear model with true values of $\mathbf{v_0}$ as the outcome and $\mathbf{u}$ as the predictor. Because the reference model represents the ideal case where we have perfect estimation of the DDM parameters, it serves as a reference for the best model we can expect from the samples. Figure \ref{sim1_1}a shows the comparison between the reference model, RegDDM and the two-step approach over the estimated effect of $\mathbf{u}$.
Compared to the reference model, the two-step approach is biased towards the null hypothesis. As the number of trials per subject increases, such bias gradually decreases. However, such bias is still non-negligible even under 150 trials per subject, which is commonly seen in practice. In contrast, no obvious bias from RegDDM is detected even with a smaller number of trials. 

We also compared the mean squared error (MSE) of estimated DDM parameters between the RegDDM and two-step approach (Figure \ref{sim1_1}b). Because RegDDM utilized information from $\mathbf{u}$ to estimate $\mathbf{v_0}$, its estimation is slightly better than two-step approach where $\mathbf{v_0}$ is only estimated using trial-level data. Such improvement shrinks as the number of trials increases, and becomes almost negligible when $n=150$. 

\begin{figure}[h]
    \centering
    \includegraphics[width=0.9\linewidth]{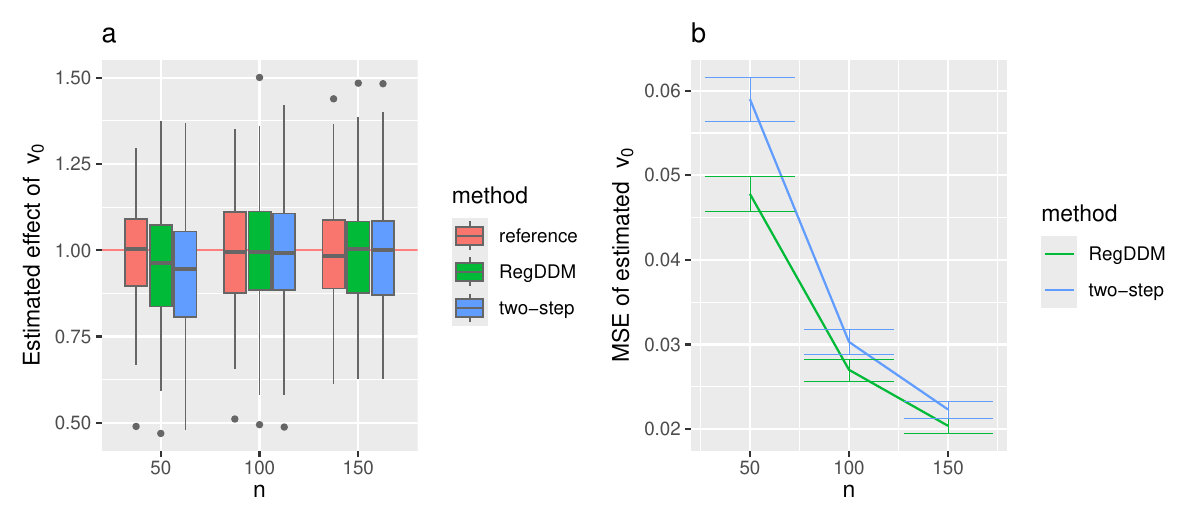}
    \caption{Comparison between methods when $v_0$ is treated as the predictor. (a) Comparison between reference model, two-step approach and RegDDM in estimating the effect of $v_0$ on $y$. Red line represents true effect. (b) MSE of estimated $v_0$ between first and second RegDDM model.}
    \label{sim1_2}
\end{figure}

Alternatively, we compared the performance between two methods when DDM parameters are treated as predictors. The simulation setup is the same as the previous simulation, except that
\begin{align*}
v_{0,i} &\sim N(1.5, 0.5^2)\\
y_i & \sim N (v_{0,i}, 0.5^2) 
\end{align*}
Then, we performed the same analysis using Equation \ref{eq9} with RegDDM, two-step approach and reference model respectively. The results were shown in Figure \ref{sim1_2}.
\begin{equation}
    y_{i} \sim N(\beta_0+\beta_{v,0}v_{0,i},\sigma^2)
    \label{eq9}
\end{equation}
When DDM parameters are treated as predictors and the number of trials per subject is low ($n = 50$), both RegDDM and the two-step approach appear to be slightly biased toward the null hypothesis, with RegDDM having slightly better performance (Figure \ref{sim1_2}a). In other cases, the bias of RegDDM and the two-step approach are similar, both showing only subtle difference compared to the reference model. The MSE of estimated DDM parameters shows trends similar to the previous result, with a descending difference between two different methods (Figure \ref{sim1_2}b). 

\subsection[Simulation2]{RegDDM model performance under different sample sizes}

To properly determine the number of trials and subjects needed, we performed the second set of simulations to test our model performance under different number of subjects ($N$), number of trials for each subject ($n$) and number of trial-level variables ($q$). Table \ref{sim2_description} shows an overview of data generation process and RegDDM model used under different number of trial-level variables. The full detail of the generation process can be found at Appendix \ref{Appendix2}.
\begin{table}[h]
\centering
\begin{tabular}{ccc}
\hline
q & Simulation setup & RegDDM model \\
\hline
0 & \makecell[c]{$v_{i,j} = v_{0,i}$} &  list(v $\sim$ 1, y $\sim$ v\_0) \\
1 & \makecell[c]{$v_{i,j} = v_{0,i} + x_{1,j}v_{x_1,i}$} & list(v $\sim$ x1, y $\sim$ v\_0 + v\_x1)\\
2 & \makecell[c]{$v_{i,j} = v_{0,i} + x_{1,j}v_{x_1,i} + x_{2,j}v_{x_1,i}$} & list(v $\sim$ x1 + x2, y $\sim$ v\_0 + v\_x1 + v\_x2)\\
\hline
\end{tabular}
\caption{Data generation setup and model used in second simulation under different number of trial-level variables (q)}
\label{sim2_description}
\end{table}

\begin{figure}[h]
    \centering
    \includegraphics[width=1\linewidth]{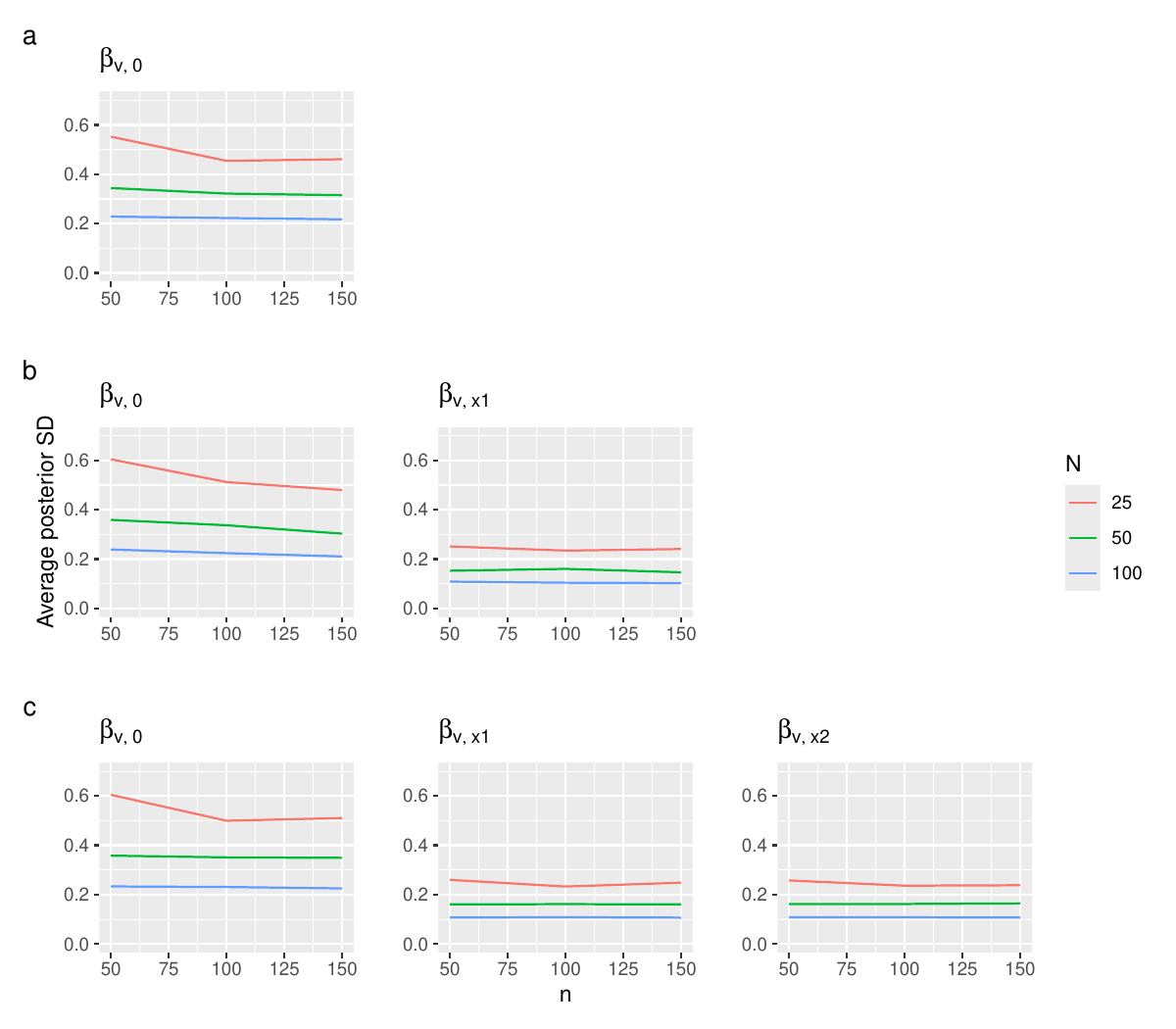}
    \caption{Posterior standard deviation of regression coefficients under different number of subjects (N) and number of trials per subject (n), when number of trial-level variable is 0 (a), 1 (b) or 2 (c).}
    \label{sim2_1}
\end{figure}

For each combination of experiment condition, 30 simulations were performed. The average posterior SD of regression coefficients are shown in Figure \ref{sim2_1}, which measures the precision of posterior estimates. It is clear that as the number of subjects increases, the posterior SD of regression coefficients became smaller. The number of trials per subject has less effect on such estimation, especially when the number of subjects is higher. Even though the estimation of DDM parameters continues to improve after 100 trials (Figure \ref{sim1_1}b, \ref{sim1_2}b), they have less effect in estimating the relationship between DDM parameters and other subject-level variables. Thus, to improve the statistical power of such analysis, the better way is to increase the number of subjects instead of using very high number of trials per subject.

The number of trial-level variables also has less influence than the number of subjects. In simulations where no trial-level variables are included (Figure \ref{sim2_1}a), the posterior standard deviation of baseline drift rate's effect on the outcome is only slightly smaller than more complex models with one or two trial-level variables (Figure \ref{sim2_1}b, c). As the number of subjects and the number of trials per subject increase, such difference becomes negligible. This observation is also found by comparing the posterior standard deviation of $\beta_{v,{x1}}$ between experiments with 1 or 2 variables. 

\begin{figure}[h]
    \centering
    \includegraphics[width=0.5\linewidth]{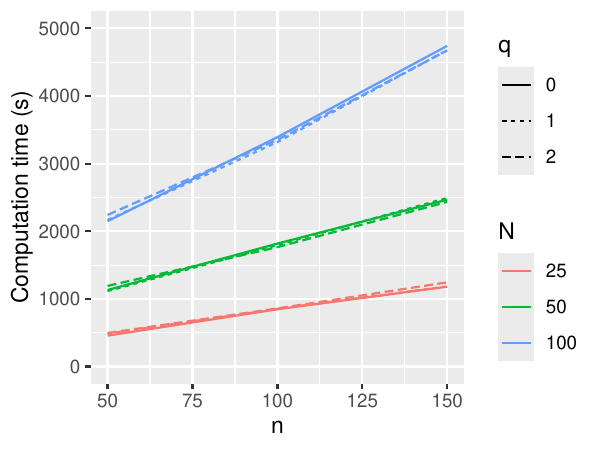}
    \caption{Computation time under different number of trials per subject (n), number of trial-level variables included (q) and  number of subjects (N)}
    \label{sim2_2}
\end{figure}

Then, we evaluated the speed of the algorithm under these conditions (Figure \ref{sim2_2}). In general, the computation time increases linearly with the number of trials and the number of subjects. Increasing the number of trial-level variables to 1 or 2 does not influence the computation time. Based on our simulation, RegDDM can apply to larger datasets with more than a thousand subjects and still converge properly within reasonable time. 

\section{Advanced topics}\label{Additional} 
\subsection[complex]{Generalized linear regression with RegDDM}
For illustration purposes, we focused on linear models in this tutorial. However, RegDDM can be expanded to the generalized linear model (GLM) if the outcome is not a DDM parameter. Currently, RegDDM supports Gaussian, Bernoulli, and Poisson family distributions. The link functions are the respective canonical links of these distributions. To fit GLM with RegDDM, the family distribution needs to be specified. For example, the following code fits a logistic regression with baseline drift rate as the predictor and gender as the outcome.
\begin{lstlisting}[basicstyle=\fontsize{8}{10}\selectfont\ttfamily]
data(regddm_data)
model = list(gender ~ v_0)
regddm(regddm_data$data1, regddm_data$data2, model, family = 'bernoulli')
\end{lstlisting}

\subsection[complex]{Multiple DDM parameters}
In addition, it is possible to model two or more DDM parameters simultaneously, but only one regression formula can be included.
\begin{lstlisting}[basicstyle=\fontsize{8}{10}\selectfont\ttfamily]
model = list(
  a ~ x1,
  v ~ x2,
  y ~ a_0 + a_x1 + v_0 + v_x2
)
\end{lstlisting}

Due to the flexibility and complexity of DDM and thus RegDDM, the user can easily build extremely complex models. However, without sufficient sample sizes, such practices often make the interpretation of the results harder, greatly reduce the statistical power, and lead to difficulties in model convergence.

\subsection[Missing data]{Handling missing data}
One advantage of Bayesian statistics is the ability to directly incorporate missing data into the rest of the model, instead of imputing the value or removing the subject \cite{merkle2011comparison}. As a Bayesian hierarchical model, RegDDM supports missing data in its subject-level covariates. For continuous variables, we assume the unknown data to follow a normal distribution. The mean and standard deviation of the normal distribution is estimated from the known parts of the variable following the same distribution:
\begin{align*}
    \mathbf{c} &= \boldsymbol c_{mis} \oplus \boldsymbol c_{obs}\\
    c_i &\sim N(\mu_c, \sigma_c^2)
\end{align*}

For factor variables, due to the nature of the Hamiltonian Monte Carlo algorithm RStan uses, missing values cannot be modeled as binary parameters. In this case, RegDDM will treat NA as a different factor level, and create dummy variables for each level including NA.

The current implementation of RegDDM does not support missing data on the trial level for computational efficiency. Users should remove rows with missing information before using RegDDM. In practice, out of many trials, we observe only several trials missing, and those few generally do not influence the posterior estimates. When a subject has missing values in a large proportion of the trials, it is recommended to examine the subject’s compliance with the tasks. With proper reasons, we could remove the subject from the analysis. 

\section{Discussion}\label{Discussion and future directions}

The DDM is a commonly used tool to study the binary decision-making process in cognitive neuroscience and related clinical research. However, in practice, integrated modeling remains challenging due to the limited accessibility to fully Bayesian approach for typical users. The R Package RegDDM implements an integrated Bayesian hierarchical model that performs generalized linear regression on DDM parameters. Compared to the conventional two-step approach, RegDDM provides a unified modeling framework, enables more robust and unbiased estimation of DDM parameters, and demonstrates superior performance under various conditions.

Our simulations illustrated the balance between the number of subjects and the number of trials per subject. Previous studies have examined this topic by assessing the estimation accuracy of DDM parameters under different numbers of trials \cite{lerche2017many}. Our second simulation showed that although the accuracy of posterior estimates continues to increase with even 150 trials per subject, the marginal benefit to downstream regression analysis between DDM parameters and other variables of interest.  Further, the number of trial-level variables that control task condition appears to have limited influence on the required number of trials. 

The current version of RegDDM is based on the four-parameter DDM, due to its popularity and the lower number of trials required. It performs robustly under mild to moderate contamination. However, under high contamination, the estimated effect of DDM parameters on the outcome may become biased. To address this limitation, future versions will include support for more complex models, such as the seven-parameter DDM \cite{ratcliff2004comparison}. 

In addition, the speed of the MCMC algorithm currently constrains the computational efficiency of RegDDM. Although the computation cost increases linearly with sample size as demonstrated in numerical simulations, the scalability of RegDDM must be improvement to handle large datasets, such as the Adolescent Brain Cognitive Development (ABCD) study \cite{casey2018adolescent, mads2023computational}. A promising future direction is to implement variational inference \cite{blei2017variational}, which requires further development and validation.

\section{Appendix}\label{Appendix}
\subsection[Appendix1]{Appendix 1: Data generatinon for Simulation 1}\label{Appendix1}
For each experiment, we generated a simulated dataset of $N = 50$ subjects with each subject having either $n = 50$, $100$ or $150$ trials. There are no trial-level variables that influence the decision process. Thus, for subject $i$ and trial $j$,
\begin{align*}
a_{i,j} &= a_{0,i}\\
t_{i,j} &= t_{0,i}\\
z_{i,j} &= z_{0,i}\\
v_{i,j} &= v_{0,i}
\end{align*}

First, we want to compare the two methods when DDM parameters are treated as the outcome. The subject-level data is generated under the following setup:
\begin{align*}
u_i &\sim N(0, 0.5^2)\\
a_{0,i} &\sim Unif(1, 3)\\
t_{0,i} &\sim Unif(0.2, 0.5)\\
z_{0,i} &\sim Unif(0.4, 0.6)\\
v_{0,i} &\sim N(1.5 + u_i, 0.5^2)\\
\end{align*}

Then, when DDM parameters are treated as predictors, the following setup was used:
\begin{align*}
v_{0,i} &\sim N(1.5, 0.5^2)\\
a_{0,i} &\sim Unif(1, 3)\\
t_{0,i} &\sim Unif(0.2, 0.5)\\
z_{0,i} &\sim Unif(0.4, 0.6)\\
y_{0,i} &\sim N(v_{0,i}, 0.5^2)\\
\end{align*}

\subsection[Appendix2]{Appendix 2: Data generatinon for Simulation 2}\label{Appendix2}

{
\allowdisplaybreaks
\begin{align*}
&a_{0,i} \sim Unif(1, 3)\\
&t_{0,i} \sim Unif(0.2, 0.5)\\
&z_{0,i} \sim Unif(0.4, 0.6)\\
&v_{0,i} \sim N(1.5, 0.5^2)\\
&a_{i,j} = a_{0,i}\\
&t_{i,j} = t_{0,i}\\
&z_{i,j} = z_{0,i}\\
&v_{i,j} =
    \begin{cases}
        v_{0,i}, & \text{if $q$ = 0}\\
        v_{0,i} + x_{1,j}v_{x_1,i}, & \text{if $q$ = 1}\\
        v_{0,i} + x_{1,j}v_{x_1,i} + x_{2,j}v_{x_1,i}, & \text{if $q$ = 2}
    \end{cases}\\
&x_1, x_2 \sim  Unif(-\sqrt{3},\sqrt{3})\\
&v_{x_1,i}, v_{x_2,i} \sim  N(0,1)\\
&y_i \sim N(0,1)
\end{align*}
}

\bmsection*{Author contributions}

This is an author contribution text. This is an author contribution text. This is an author contribution text. This is an author contribution text. This is an author contribution text.

\bmsection*{Acknowledgments}
The work is supported by the National Institutes of Health/National Institute on Aging (NIH/NIA; grant numbers R01 AG038465, R01 AG026158, and R01 AG062578).

We are grateful to Margaret Gacheru for her helpful comments and suggestions during the revision of this paper.

\bmsection*{Financial disclosure}

None reported.

\bmsection*{Conflict of interest}

The authors declare no potential conflict of interests.

\bibliography{wileyNJD-AMA}

\bmsection*{Supporting information}

Additional supporting information may be found in the
online version of the article at the publisher’s website.

\bmsection*{Author Biography}

\begin{biography}{\includegraphics[width=76pt,height=76pt,draft]{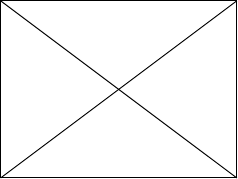}}{
{\textbf{Author Name.} Please check with the journal's author guidelines whether
author biographies are required. They are usually only included for
review-type articles, and typically require photos and brief
biographies for each author.}}
\end{biography}

\end{document}